\documentclass[aps,prb,epsf,twocolumn,showpacs]{revtex4-2}

\usepackage{CJK,amsmath,amssymb,graphics,epsfig,epstopdf,color,verbatim,ulem,braket,tabularx,float}
\usepackage[colorlinks,linkcolor=blue,citecolor=blue,urlcolor=blue]{hyperref}

\begin{document}
\begin{CJK*}{GBK}{song}
\title{\mbox{Magnetization Plateaus in the Two-dimensional $S$ = 1/2 Heisenberg } 
\mbox{Model with a 3$\times$3 Checkerboard Structure}}

\author{Xuyang Liang and Dao-Xin Yao}
\email{yaodaox@mail.sysu.edu.cn}
\affiliation{Guangdong Provincial Key Laboratory of Magnetoelectric Physics and Devices, State Key Laboratory of Optoelectronic Materials and Technologies, Center for Neutron Science and Technology, School of Physics, Sun Yat-Sen University, Guangzhou, 510275, China}

\begin{abstract}
We investigate the $S$=1/2 antiferromagnetic Heisenberg model with a 3$\times$3 checkerboard lattice structure in a longitudinal magnetic field. By using the stochastic series expansion quantum Monte Carlo (SSE-QMC) method, we obtain the properties of the non-plateau XY phase, 1/9, 3/9, 5/9, 7/9 magnetization plateau phases, and fully polarized phase. Then, we determine the precise phase transition critical points belonging to the 3D XY universality class through finite-size scaling. Moreover, we study the longitudinal and transverse dynamic spin structure factors of this model in different phases. For the non-plateau XY phase, the energy spectra present a gap between the low-energy gapless branch and the high-energy part under the competition of magnetic field and interaction. The gapless branch can be described by the spin wave theory in the canted antiferromagnetic phase of the effective ``block spin'' model.  In the magnetization plateau phase, we identify that the excitation arises from localized disturbances within the sublattice, which are capable of spreading in momentum space. This study offers theoretical insights and interpretations for the characteristics of the ground state and the inelastic neutron scattering spectrum in two-dimensional quantum magnetic materials. Specifically, it focuses on materials with a checkerboard unit cell structure and odd-spin configurations under the influence of a longitudinal magnetic field.

\end{abstract}


\date{\today}
\maketitle
\end{CJK*}

\section{Introduction}

The $S$ = 1/2 Heisenberg antiferromagnetic model has consistently garnered attention in the realms of condensed matter physics, due to its crucial significance in the investigation of cuprate superconductors~\cite{cupratesuperconductors1, cupratesuperconductors2, cupratesuperconductors3, cupratesuperconductors4}, and quantum magnetic materials ~\cite{magneticmaterials1,magneticmaterials2}. Among that, the magnetization plateau is a quantum phenomenon of the antiferromagnetic Heisenberg model, which has been extensively reported in both theory and experiments. In the one-dimensional Haldane chain~\cite{haldane1983continuum,haldane1983nonlinear}, the zero magnetization plateau appears until a finite magnetic field closes the gap between the singlet of the ground state and the triplet state of the first excited state. Notably, all plateaus conform to the Oshikawa Yamanaka-Affleck (OYA) criterion $(n(s-m)=integer)$ in one-dimensional spin system~\cite{oshikawa}, where $n$ represents the number of sites in each unit cell, $s$ is the magnitude of spin and $m$ is the magnetization per spin. Moreover, the magnetization plateau induced by quantum fluctuations has become a very active research field in geometric frustration systems~\cite{TriangularMagnetizationPlateau1,TriangularMagnetizationPlateau2,TriangularMagnetizationPlateau3,TriangularMagnetizationPlateau4,J1-J2TriangularMagnetizationPlateau1,J1-J2TriangularMagnetizationPlateau2,J1-J2TriangularMagnetizationPlateau3,J1-J2TriangularMagnetizationPlateau4,checkerboardMagnetizationPlateau1, checkerboardMagnetizationPlateau2}.

Besides the properties of the ground state, we are interested in the dynamic properties of magnetic systems. In recent years, inelastic neutron scattering(INS) experiments have revealed many novel dynamic properties. The high-energy continuum appears near ($\pi$,0) in the INS experiments of La$_2$CuO$_4$ and Cu(DCOO)$_2$$\cdot$4D$_{2}$O, which cannot be described by the linear spin wave of square lattice~\cite{LSWT1, LSWT2}. Furthermore, the magnetic field induced spontaneous magnon decays are observed in the compound Ba${_2}$MnGe${_2}$O${_7}$~\cite{masuda2010}. On the other hand, the numerical simulations also provide valuable results on exploring the dynamical properties. For example, the method of combining stochastic analytic continuation(SAC) with QMC can effectively reveal the dynamic properties of spin liquids and fractionalization at a deconfined quantum critical point~\cite{spinliquild1,DQCP1}.

In two-dimensional (2D) systems, certain materials show checkerboard structures as evidenced by experimental studies, such as Bi$_2$Sr$_2$CaCu$_2$O$_{8+\delta}$ and Ca$_{2-x}$Na$_x$CuO$_2$Cl$_2$ ~\cite{material1,material2}. In previous work, the magnetic excitation mechanism of the checkerboard model has been investigated~\cite{checkerboardran,checkerboardxu}. Notably, the $3\times3$ checkerboard model shows the excitation of spin wave originating from inter-sublattice, due to each $3\times3$ plaquette having the odd number of spins. Furthermore, the one-dimensional trimer chains are a simplification of this model. It is found that the two-spinon continuum is described by the novel intermediate-energy and high-energy localized excitations are termed as the ``doublon'' and ``quarton'' for weak inter-trimer interaction $J_2$, respectively.~\cite{trimer1}. These excitations have been confirmed to exist in the experimental material Na$_2$Cu$_3$Ge$_4$O$_{12}$~\cite{trimer2}. Futhermore, such excitations are observed in the 2D theoretical model proposed based on Ba$_{4}$Ir$_{3}$O$_{10}$ and CaNi$_{3}$(P$_{2}$O$_{7}$)$_{2}$ with trimerized structure~\cite{Ba4Ir3O101,Ba4Ir3O102,Ba4Ir3O103,Ba4Ir3O104,CaNi3P2O72,chang2024magnon}.

Recently, the one-dimensional trimer chains emerges the XY phase and 1/3 magnetization plateau phase under the competition between the magnetic field and interaction, and propagates the ``doublon'' and ``quarton'' in a longitudinal magnetic field.~\cite{cheng2024quantum}. For the 2D system, the $3\times3$ checkerboard model may induce more magnetization plateaus and different fractional spin excitations in a magnetic field. In this paper, we investigate the properties of the ground state and spin dynamic of the $S$ = 1/2 Heisenberg model on a 3$\times$3 checkerboard lattice in a longitudinal magnetic field by using the SSE-QMC and SAC. 

The rest of the paper is organized as follows. In Sec.~\ref{Sec:MODEL, METHOD, AND DEFINITIONS}, we introduce the Hamiltonian of this model and the QMC-SAC method. In Sec.~\ref{Sec:NUMERICAL RESULTS}, we study the critical points of phase transition based on the theory of finite-size scaling and present the results of the dynamic structural factors of the 3$\times$3 checkerboard model in different magnetic fields. In Sec.~\ref{Sec:Discussion}, we explain the magnetic excitation mechanism of the magnetization plateau based on perturbation theory and discuss the excitation of the high-energy mode. The Sec.~\ref{Sec:CONCLUSION} is the summary of this article.

\begin{figure}
    \centering
    \includegraphics[width=1.0\linewidth]{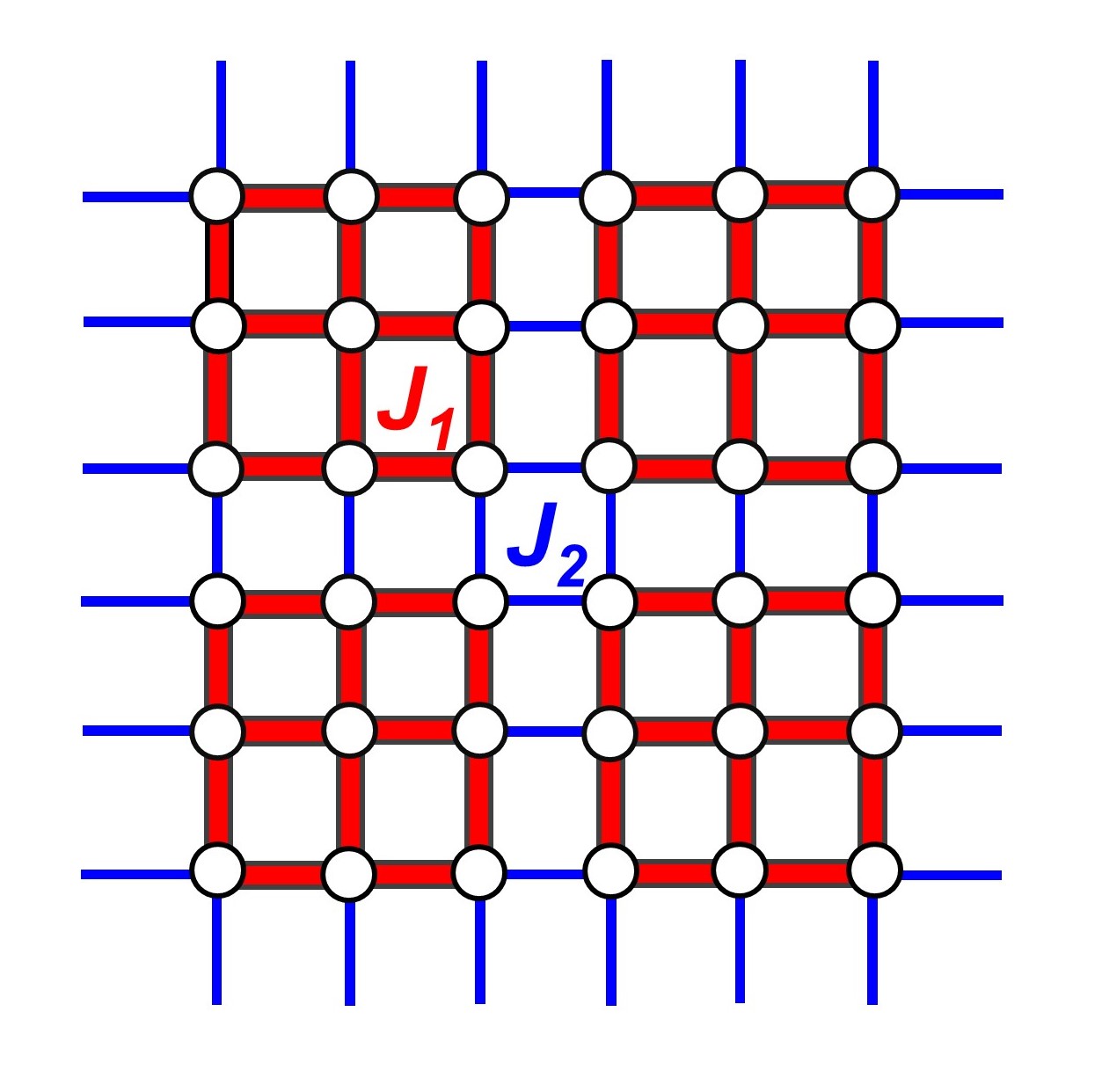}
    \caption{Structure of 3$\times$3 checkerboard lattice. The intra-sublattice interaction $J_1$ and the inter-sublattice interaction $J_2$ are represented by thick red and thin blue bonds, respectively.}
    \label{fig:checkerboard}
\end{figure}

\section{MODEL, METHOD, AND DEFINITIONS}
\label{Sec:MODEL, METHOD, AND DEFINITIONS}
\subsection{Model}
\label{Sec:model}
The Hamiltonian of the $S$ = 1/2 antiferromagnetic Heisenberg model on a 3$\times$3 checkerboard lattice in a longitudinal magnetic field can be written as
\begin{equation}
\begin{array}{l}
H=J_1\sum\limits_{\left \langle i,j\right \rangle}{{{S}}_{i} \cdot {{S}}_{j}}+J_2\sum\limits_{\left \langle i,j \right \rangle'}{{S}}_{i} \cdot {{S}}_{j}-h{\textstyle \sum\limits_{i}}{{S}}_{i}^{z},
\end{array}
\label{Eq:Hmlt}
\end{equation}
where $S_i$ denotes the spin-1/2 operator on each site i; $\left \langle i,j\right \rangle$  and $\left \langle i,j \right \rangle'$represent the nearest-neighbor sites. The strong coupling $J_1$ and weak coupling $J_2$ correspond to the thick red and the thin blue bonds respectively, which represent antiferromagnetic nearest-neighbor interactions, and the longitudinal magnetic field is denoted by $h$ in Fig.~\ref{fig:checkerboard}. We refer to each 3$\times$3 plaquette as a sublattice and define the coupling ratio as g=$J_2/J_1$ in this paper. The value of $J_1$ is fixed to 1 as the energy unit, where the value of $J_2$ corresponds to g. We are interested in the whole range of coupling ratios g from 0 to 1, where the system evolves between the isolated sublattices(g=0) to antiferromagnetic Heisenberg square lattice with uniform interactions(g=1).

\subsection{Quantum Monte Carlo method}
\label{Sec:QMC}
We use the SSE-QMC simulation method to study the phase diagram and spectral functions of the 3$\times$3 checkerboard model in this work~\cite{directloop1}. Via SSE-QMC, we can obtain imaginary-time correlation functions. Due to the influence of the magnetic field, the symmetry of SU(2) is broken. To comprehensively reveal the dynamic properties, the longitudinal and transverse imaginary-time correlation functions are measured, which are defined as 

\begin{equation}
\begin{array}{l}
G_{q}^{zz}(\tau) =\frac{1}{N} \sum\limits_{ i,j }e^{-iq(r_i-r_j)}  (\left \langle S_{i}^{z}(\tau)S_{j}^{z}(0)\right \rangle - \left \langle S_{i}^{z}(\tau)\right \rangle \left \langle S_{j}^{z}(0)\right \rangle),
\end{array}
\label{Eq:G(tauzz)}
\end{equation}

\begin{equation}
\begin{array}{l}
G_{q}^{\pm }(\tau) =\frac{1}{4N}\sum\limits_{ i,j }e^{-iq(r_i-r_j)}  (\left \langle S_{i}^{+}(\tau)S_{j}^{-}(0)+S_{i}^{-}(\tau)S_{j}^{+}(0)\right \rangle, 
\end{array}
\label{Eq:G(taupm)}
\end{equation}
where $r_i$ is the spatial position of the site i, the dynamic spin structure factor can be obtained by 

\begin{equation}
\begin{array}{l}
G_q\left(\tau\right)=\int_{-\infty}^{\infty}{d\omega \mathcal{S}(q,\omega)e^{-\tau\omega}}.
\end{array}
\label{Eq:SAC}
\end{equation}
In the SAC, the spectral function $\mathcal{S}(q, \omega)$ is parameterized by a large number of $\delta$ functions, and then the imaginary-time correlation functions $G_q(\tau)$ will be fitted through Monte Carlo simulation~\cite{SAC1,SAC2}.

\section{NUMERICAL RESULTS}
\label{Sec:NUMERICAL RESULTS}
Here, we mainly study the 3$\times$3 checkerboard lattice model with periodic boundary conditions under the influence of a longitudinal magnetic field.

\subsection{Quantum phase transitions}
 To explore ground-state properties, we set up the inverse temperature $\beta=4L$ in QMC simulation unless otherwise specified. First, we study the phase diagram of the $S =1/2$ Heisenberg model on the 3$\times$3 checkerboard square lattice in a longitudinal magnetic field. To investigate quantum phase transitions through QMC simulations, we employ spin stiffness $\rho_s$ to probe the quantum critical points. Spin stiffness $\rho_s$ is defined by the free energy $f$ and the twist angle $\phi$,

\begin{equation}
\begin{array}{l}
\rho_s=\frac{1}{N}\frac{\partial^2f}{\partial\phi^2},
\end{array}
\label{Eq:stiffness}
\end{equation}
in the SSE simulations, the $\alpha$(x or y) direction of spin stiffness can be calculated from

\begin{figure}[t]
    \centering
    \includegraphics[width=1\linewidth]{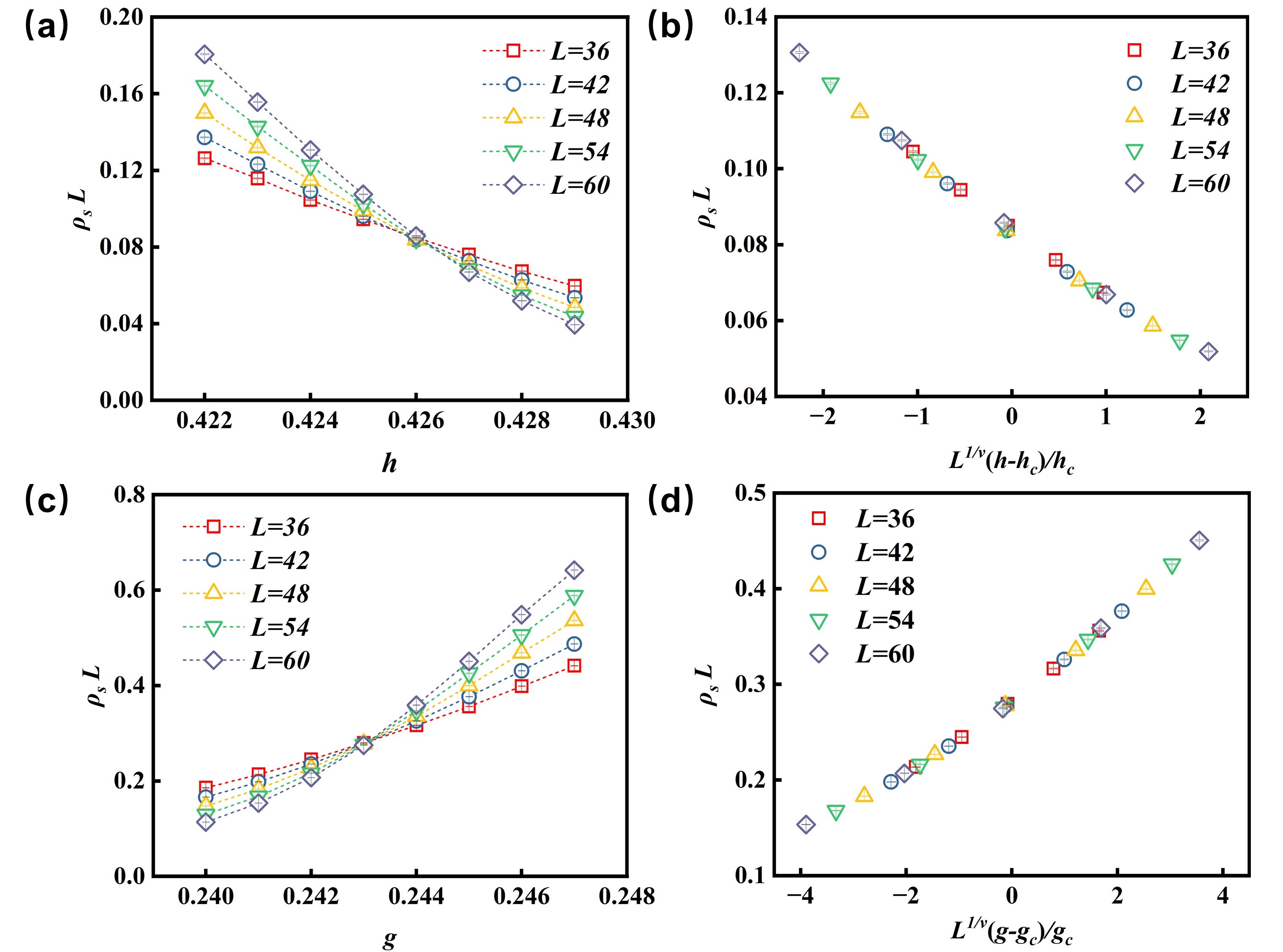}
    \caption{(a)Spin stiffness multiplied by $L$ at fixed g=0.2 versus the magnetic field $h$, and (c) spin stiffness multiplied by $L$ at fixed $h$=0.575 versus the coupling ratio g. Panels (b) and (d) show the data collapses of spin stiffness multiplied by L, where the best fitting critical exponents $\nu$ are 0.667(5) and 0.669(2), respectively.}
    \label{fig:finite-size}
\end{figure}

\begin{equation}
\begin{array}{l}
\rho^\alpha=\frac{1}{\beta N} \left\langle(N_{\alpha}^{+}-N_{\alpha}^{-})^{2} \right \rangle, 
\end{array}
\label{Eq:stiffness2}
\end{equation}
 where $N_\alpha^+$ and $N_\alpha^-$ are actually the total number of the nearest off-diagonal operators $S_i^+S_j^-$ and $S_i^-S_j^+$ transporting spin along the positive and negative $\alpha$ direction, respectively. In fact, we only calculate $\rho_s=\frac{\left(\rho^x+\rho^y\right)}{2}$ because the 3$\times$3 checkerboard square lattice has isotropy. Meanwhile, at a quantum critical point, it should scale as

\begin{equation}
\begin{array}{l}
\rho_s\sim L^{2-d-z},  
\end{array}
\label{Eq:stiffness3}
\end{equation}
where $d$ is the dimension of the system, and $z$ is the dynamic critical exponent. In our model, we have $d$ = 2 and $z$ = 1, and the scaling becomes $\rho_s\sim~L^{-1}$. Thus, the $\rho_sL$ is size-independent at the critical point. 

To obtain the quantum critical point, we determine the critical points $h_c$ or g$_c$ as shown in Fig~\ref{fig:finite-size}(a) and (c), the clear crossing points indicate that a quantum phase transition occurs between the non-plateau XY phase and the 1/9 magnetization plateau phase.

On the basis of the theory of finite-size scaling, the spin stiffness scales as~\cite{fisher}   

\begin{equation}
\begin{array}{l}
\rho_s\left(t,L\right)L=f(tL^\frac{1}{\upsilon}),
\end{array}
\label{Eq:stiffness4}
\end{equation}
where $\nu$ is the correlation length exponent and $t=(\alpha-\alpha_c)/\alpha_c$, and the $\alpha_c$ represents $h_c$ or g$_c$. We can obtain the $\alpha_c$ and $\nu$ from a data collapse, where the values of $\nu$ are 0.667(5) and 0.669(2) as shown in Figs.~\ref{fig:finite-size}(b) and (d), respectively, indicating that the phase transition may belong to 3D XY universal class~\cite{3DXY}.

\begin{figure}
    \centering
    \includegraphics[width=1.0\linewidth]{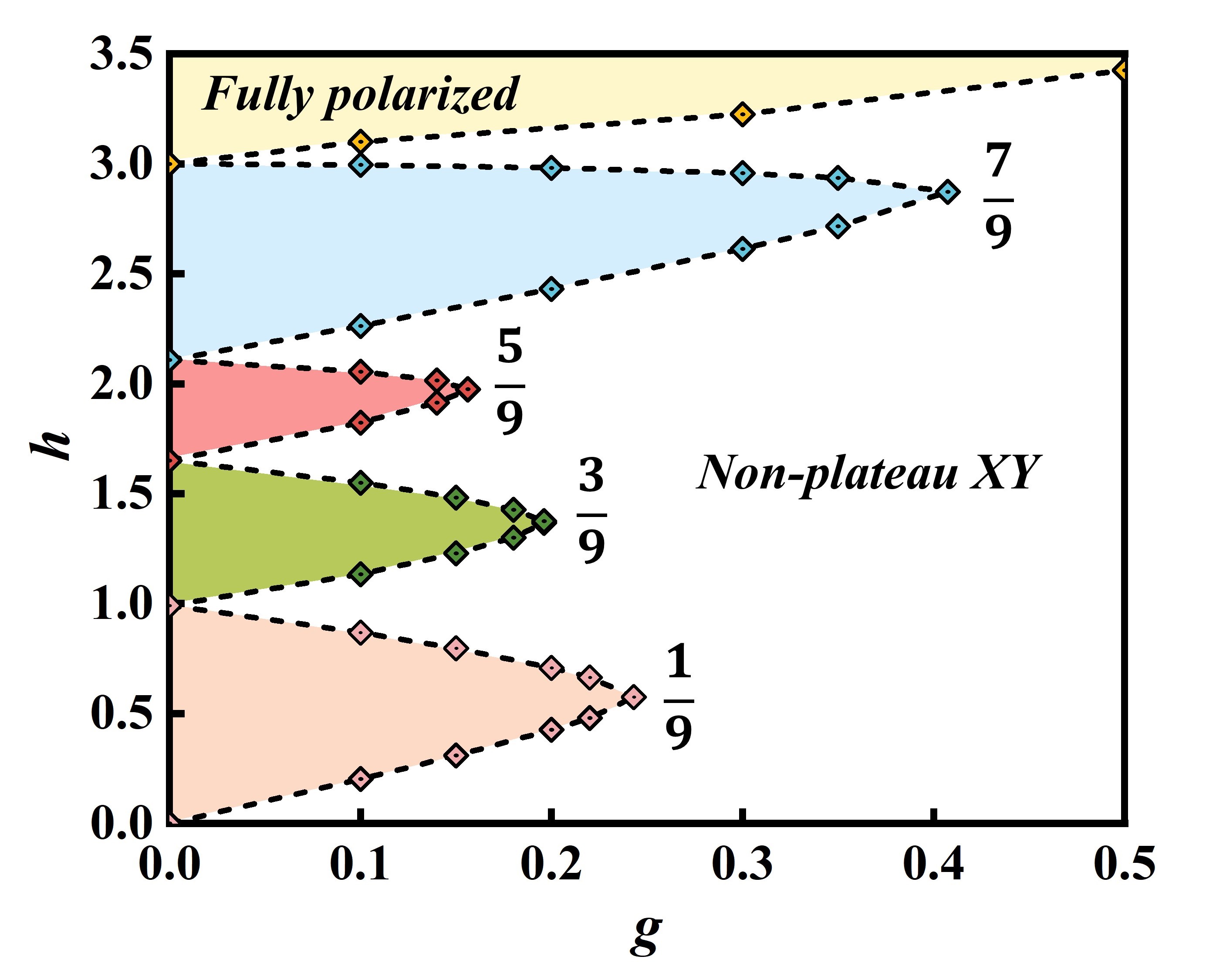}
    \caption{The phase diagram of the S = 1/2 antiferromagnetic Heisenberg model on the 3$\times$3 checkerboard lattice in a longitudinal magnetic field.}
    \label{fig:phasediagram}
\end{figure}

\begin{figure}
    \centering
    \includegraphics[width=1.0\linewidth]{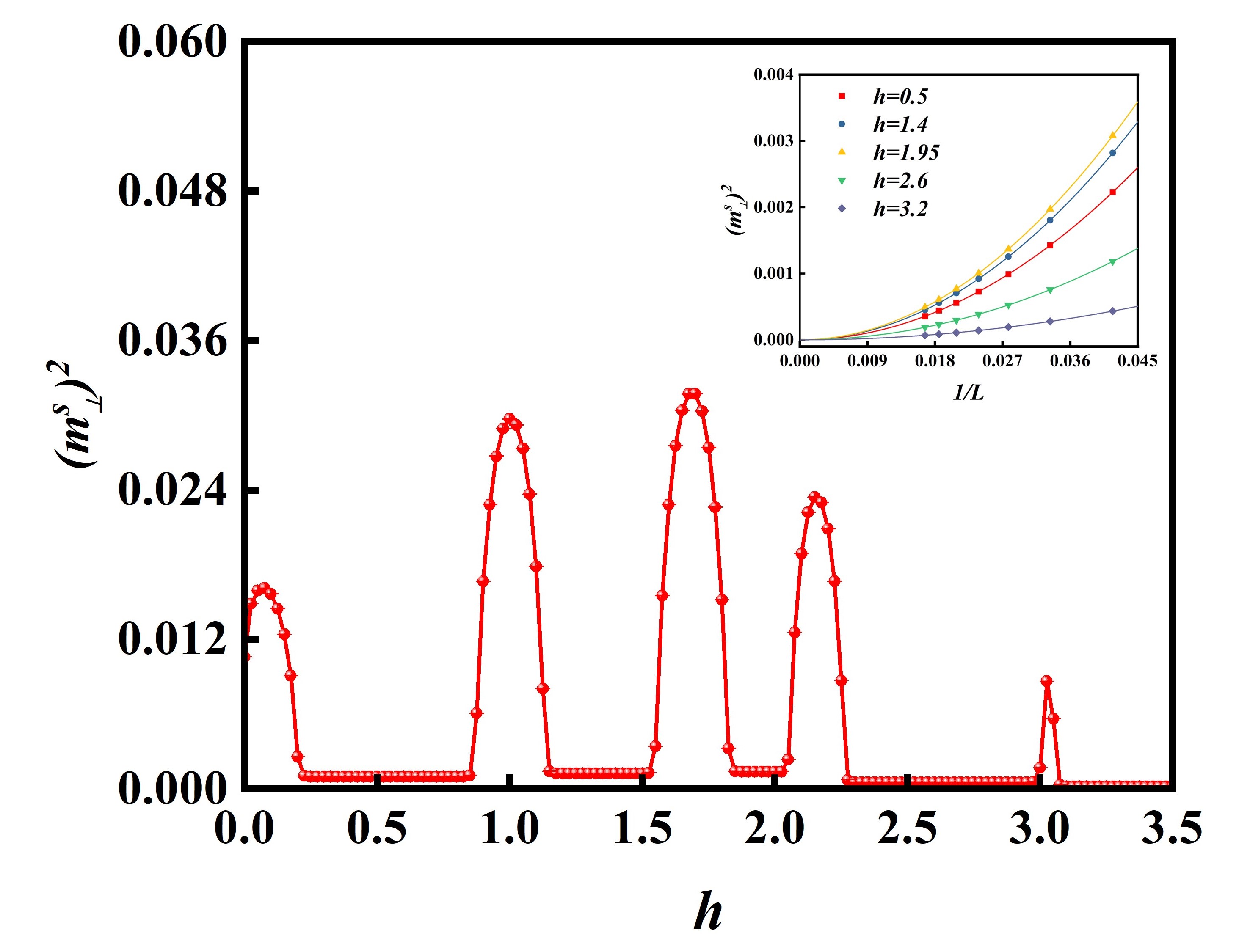}
    \caption{The transverse staggered magnetization $(m_\bot^s)^2$ for g=0.1. The inset shows the results of finite size extrapolation in different magnetization plateau phases.}
    \label{fig:ms}
\end{figure}

By using this method to calculate other values, we ultimately obtain the phase boundaries in Fig.~\ref{fig:phasediagram}. From this phase diagram, we can observe six different phases: 1/9, 3/9, 5/9, and 7/9 magnetization plateau phases, fully polarized phase, and non-plateau XY phase. To study the transverse long-range magnetic order of different phases, we define the square transverse staggered magnetization as

\begin{equation}
\begin{array}{l}
(m_\bot^s)^2=\frac{1}{4 N^{2}}(\sum\limits_{i j}(-1)^{i+j}\left\langle S_{i}^{+} S_{j}^{-}+S_{i}^{-} S_{j}^{+}\right\rangle),
\end{array}
\label{Eq:ms}
\end{equation}
where $(-1)^{i+j} = \pm 1$ represents staggered phase factors. In Fig.~\ref{fig:ms}, we show the transverse square staggered magnetization versus the magnetic field $h$ in g=0.1. In the magnetization plateau phase, the size extrapolation shows that the transverse square staggered magnetization $(m_\bot^s)^2$ is zero in the thermodynamic limit as shown in the inset of Fig.~\ref{fig:ms}.

\begin{figure*}
    \centering
    \includegraphics[width=1\linewidth]{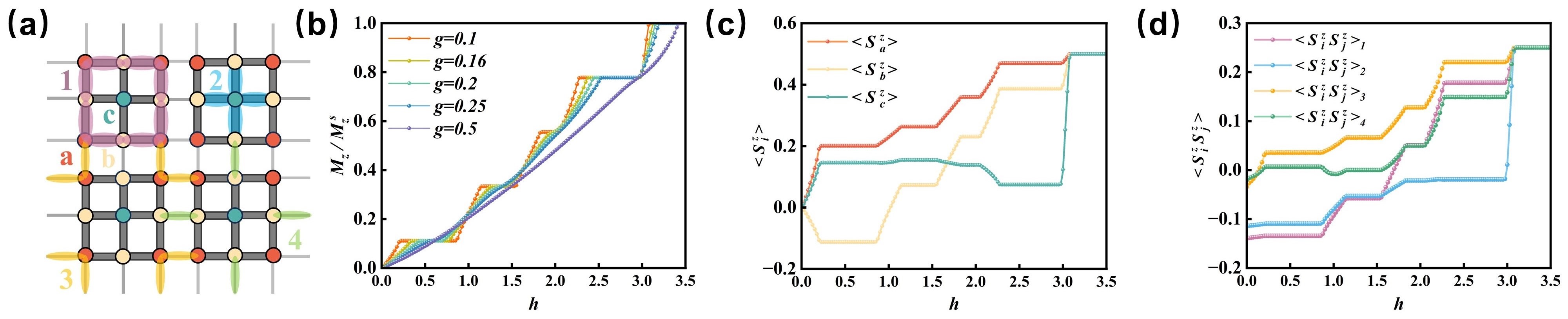}
    \caption{(a) The structure of the $3\times3$ checkerboard lattice, where we use the same color to represent the equivalent spins and spin correlations. (b) The ratio of magnetization and saturated magnetization versus the magnetic field. (c) The magnetization of spins a, b, and c with g=0.1 versus the magnetic field (d) The spin correlations $\left \langle S_{i}^{z}S_{j}^{z}\right \rangle_1$, $\left \langle S_{i}^{z}S_{j}^{z}\right \rangle_2$, $\left \langle S_{i}^{z}S_{j}^{z}\right \rangle_3$, and $\left \langle S_{i}^{z}S_{j}^{z}\right \rangle_4$ with g=0.1 versus the magnetic field.}
    \label{fig:zuhetu}
\end{figure*}

In Fig.~\ref{fig:zuhetu}(b), the magnetization curve shows the magnetization process, where the magnetization per spin is defined as $M_z=\sum_{i=1}^{N}{S_i^z/N}$, and $M_z^s$ is the saturation magnetization per spin. We find that there are 1/9, 3/9, 5/9, and 7/9 magnetization plateaus when g is small. Meanwhile, the width of the plateaus decreases as g increases. Ultimately, all the magnetization plateaus vanish.

In addition, the magnetization of each spin and the nearest-neighbor spin correlation are calculated. The equivalent spins and spin correlations are represented by the same color in Fig.~\ref{fig:zuhetu}(a), we observe that it cannot form a 1/9 magnetization plateau phase due to the magnetic field is not strong enough to open an energy gap at small $h$ as shown in Fig.~\ref{fig:zuhetu}(b). As the magnetic field increases, antiferromagnetic correlations can be slowly suppressed, leading to the formation of a 1/9 magnetization plateau phase. The four inequivalent spin correlations are as shown in Fig.~\ref{fig:zuhetu}(d), where the spin correlations $\left \langle S_{i}^{z}S_{j}^{z}\right \rangle_3$ and $\left \langle S_{i}^{z}S_{j}^{z}\right \rangle_4$ are ferromagnetic correlations, and the $\left \langle S_{i}^{z}S_{j}^{z}\right \rangle_1$ and $\left \langle S_{i}^{z}S_{j}^{z}\right \rangle_2$ are antiferromagnetic correlations in the 1/9 magnetization plateau. Moreover, It is worth noting that the magnetizations of spin and spin correlations are fixed in the magnetization plateau as shown in Figs.~\ref{fig:zuhetu}(c) and 5(d). In the 1/9 magnetization plateau phase, the magnetizations of three spins a, b, and c are 0.20095(3), -0.11243(4), 0.14593(6), reflecting that each 3$\times$3 sublattice can be regarded as an effective polarized spin 1/2. Furthermore, other details of inequivalent spin magnetizations in the corresponding magnetization plateau phase are shown in Table~\ref{tab:magnetization}, where each 3$\times$3 sublattice can be regarded as an effectively polarized spin 3/2, 5/2, and 7/2 in the 3/9, 5/9 and 7/9 magnetization plateau phases, respectively. Moreover, as the magnetic field increases, the spins a and b are further polarized until they are fully aligned. However, the nonmonotonic $h$-dependence behavior of the magnetization of spin c is observed in the regions 1.55$\le$$h$$\le$1.82 and 2.05$\le$$h$$\le$2.26, while the antiferromagnetic $\left \langle S_{i}^{z}S_{j}^{z}\right \rangle_2$ is stronger than others. Due to the magnetic field failing to suppress the strong coupling $J_1$ to completely polarize the spin c.

\begin{table}[]
\caption{The magnetization of each inequivalent spin in different magnetization plateau phases.}
\begin{tabular}{cccc}

\hline\hline
Magnetization plateau & a          & b           & c          \\ \hline
1/9($h$=0.5)            & 0.20095(3) & -0.11243(4) & 0.14593(6) \\
3/9($h$=1.35)           & 0.26341(3) & 0.07287(4)  & 0.15489(7) \\
5/9($h$=1.95)           & 0.35959(2) & 0.23077(3)  & 0.13853(7) \\
7/9($h$=2.6)            & 0.46954(1) & 0.38688(1)  & 0.07429(5) \\ \hline\hline
\label{tab:magnetization}
\end{tabular}
\end{table}

\subsection{ Excitation spectra}

In this section, we study the longitudinal and transverse dynamical structure factors $S^{zz}(q,\omega )$ and $S^{\pm}(q,\omega )$ of the $3\times3$ checkerboard model in a longitudinal magnetic field by taking g=0.7, 0.4 and 0.1. The lattice size $L$ is chosen as 48 and 36 to calculate $S^{zz}(q,\omega )$ and $S^{\pm}(q,\omega )$ respectively when g=0.7 and 0.4, and 48 for both $S^{zz}(q,\omega )$ and $S^{\pm}(q,\omega )$ when g=0.1. Inverse temperature is chosen to $\beta=L$ in QMC calculation. The results of $S^{zz}(q,\omega )$ and $S^{\pm }(q,\omega )$ are displayed along the high symmetry path $(0,0)\to  (\pi, 0)\to  (\pi, \pi)\to  (0,0) \to  (0, \pi)\to (\pi, 0)$ in the Brillouin zone. Among that, the convergence of the gapless point $(\pi,\pi)$ with the spectral weight concentrated requires a very large $\beta$~\cite{SAC1}. Consequently, we do not show it in the transverse excitation spectrum.

\begin{figure*}[t]
    \centering
    \includegraphics[width=1\linewidth]{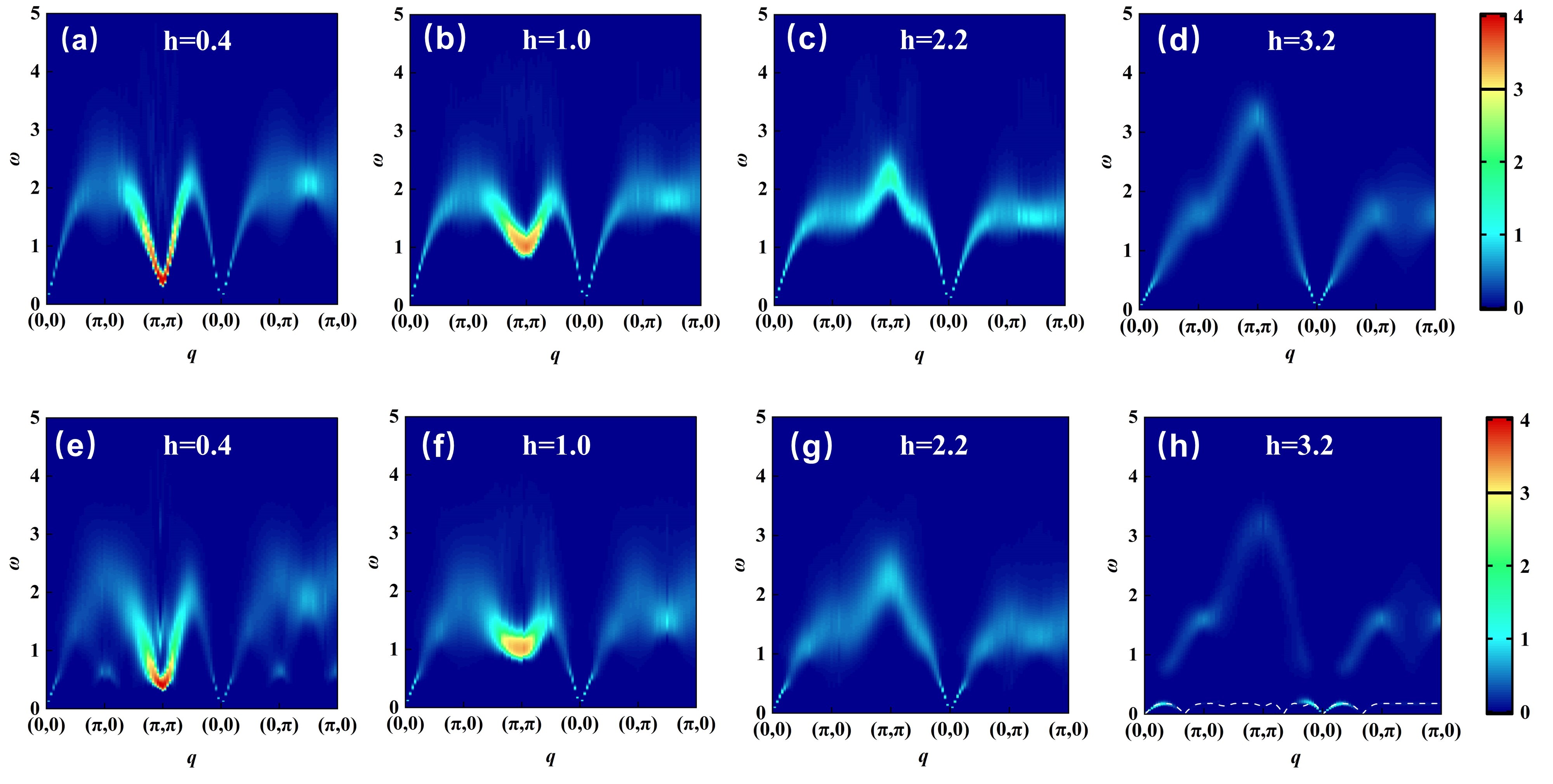}
    \caption{Dynamic spin structure factor $S^{zz}(q,\omega)$ of $3\times3$ checkerboard model obtained from QMC-SAC calculations in a fixed coupling ratio g=0.7 (a)-(d) and g=0.4 (e)-(h) with different $h$. Panels (a)-(h) are in the non-plateau XY phase. To present the results of SAC more clearly, we take the value U$_0$=3 corresponding to the black bar as the boundary. Below the boundary, the spectral function is linearly mapped to the color bar, while above the boundary, it is a logarithmic mapping, U = U$_0$ + log$_{10}$$S(q,\omega)$ - log$_{10}$U$_0$.}
    \label{fig:g=0.7and0.4zz}
\end{figure*}

\begin{figure*}[t]
    \centering
    \includegraphics[width=1\linewidth]{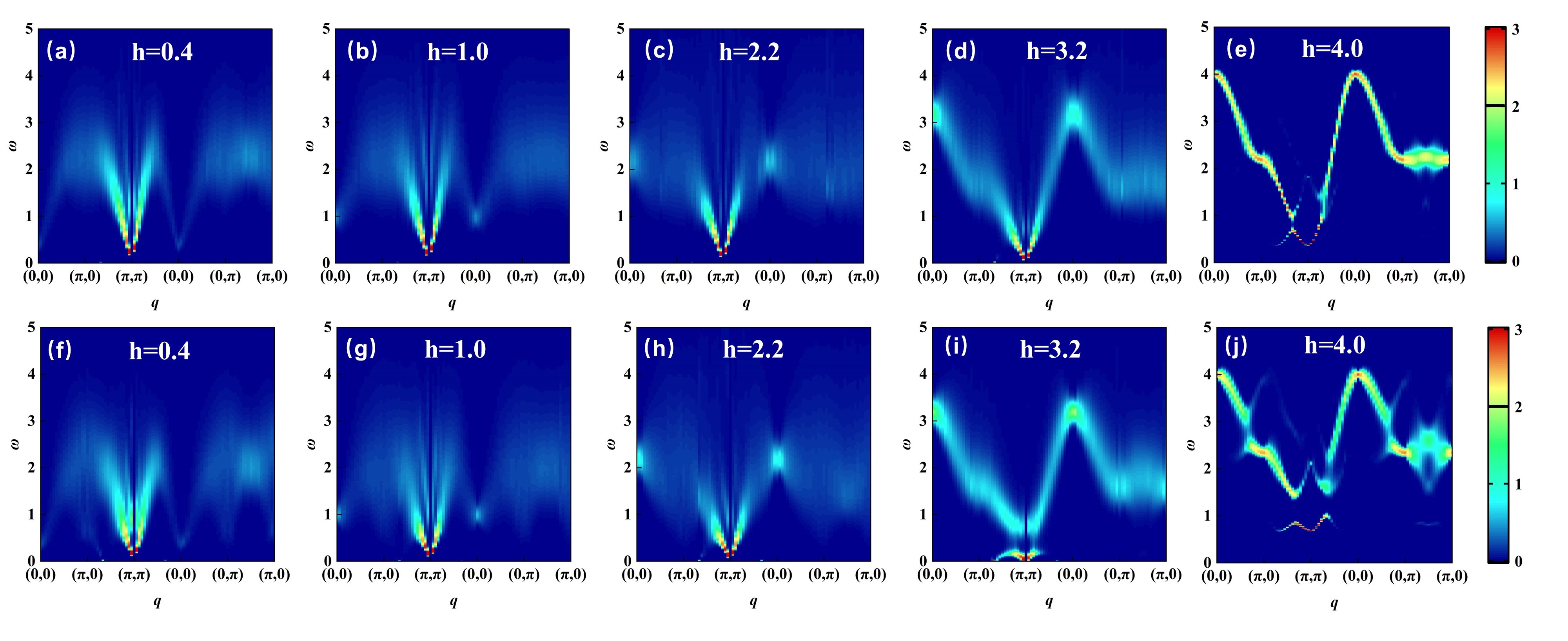}
    \caption{Dynamic spin structure factor $S^{\pm}(q,\omega)$ of $3\times3$ checkerboard model obtained from QMC-SAC calculations in a fixed coupling ratio g=0.7 (a)-(e) and g=0.4 (f)-(j) with different $h$. Panels (a)-(d) and (f)-(i) are in the non-plateau XY phase, (e) and (g) are in the fully polarized phase. We take the value U$_0$=2 corresponding to the black bar as the boundary. Below the boundary, the spectral function is linearly mapped to the color bar, while above the boundary, it is a logarithmic mapping, U = U$_0$ + log$_{10}$$S(q,\omega)$ - log$_{10}$U$_0$.}
    \label{fig:g=0.7and0.4xx}
\end{figure*}

Firstly, we investigate the results of the $3\times3$ checkerboard model $S^{zz}(q,\omega )$ in a longitudinal magnetic field. In previous work, the $3\times3$ checkerboard lattice remains in the antiferromagnetic N\'{e}el phase, which exists gapless Goldstone mode at (0,0) and $(\pi, \pi)$~\cite{checkerboardxu}. The system enters from the antiferromagnetic N\'{e}el phase to the non-plateau XY phase driven by the magnetic field. When g=0.7, we find that the structures of the spectra are similar to the square lattice in a magnetic field~\cite{magneticsquare}, there is a gap proportional to the magnetic field at $(\pi, \pi)$ and the spontaneous breaking of the $U(1)$ symmetry leads to the presence of a gapless Goldstone mode at (0,0). As the magnetic field increases, the gap at $(\pi, \pi)$ increases, and the strong magnetic field leads to the spontaneous decay of magnons. Especially, the spectrum weight tends to vanish except for the vicinity of (0,0) in the strong magnetic field as shown in Fig.~\ref{fig:g=0.7and0.4zz}(d). For the spectra of $S^{\pm}(q,\omega )$, the Goldstone theorem indicates that antiferromagnetic XY phase breaks the U(1) symmetry and contributes to the strong gapless mode at $(\pi,\pi)$. The spectrum weight is concentrated in the V-shaped structure around $(\pi, \pi)$, and the gap at (0,0) is proportional to the magnetic field as shown in Figs.~\ref{fig:g=0.7and0.4xx}(a)-(d). In the fully polarized phase, all spins are polarized, and the excitation with apparent dispersion structure mainly consists of the magnons as shown in Fig.~\ref{fig:g=0.7and0.4xx}(e).

\begin{figure}
    \centering
    \includegraphics[width=1\linewidth]{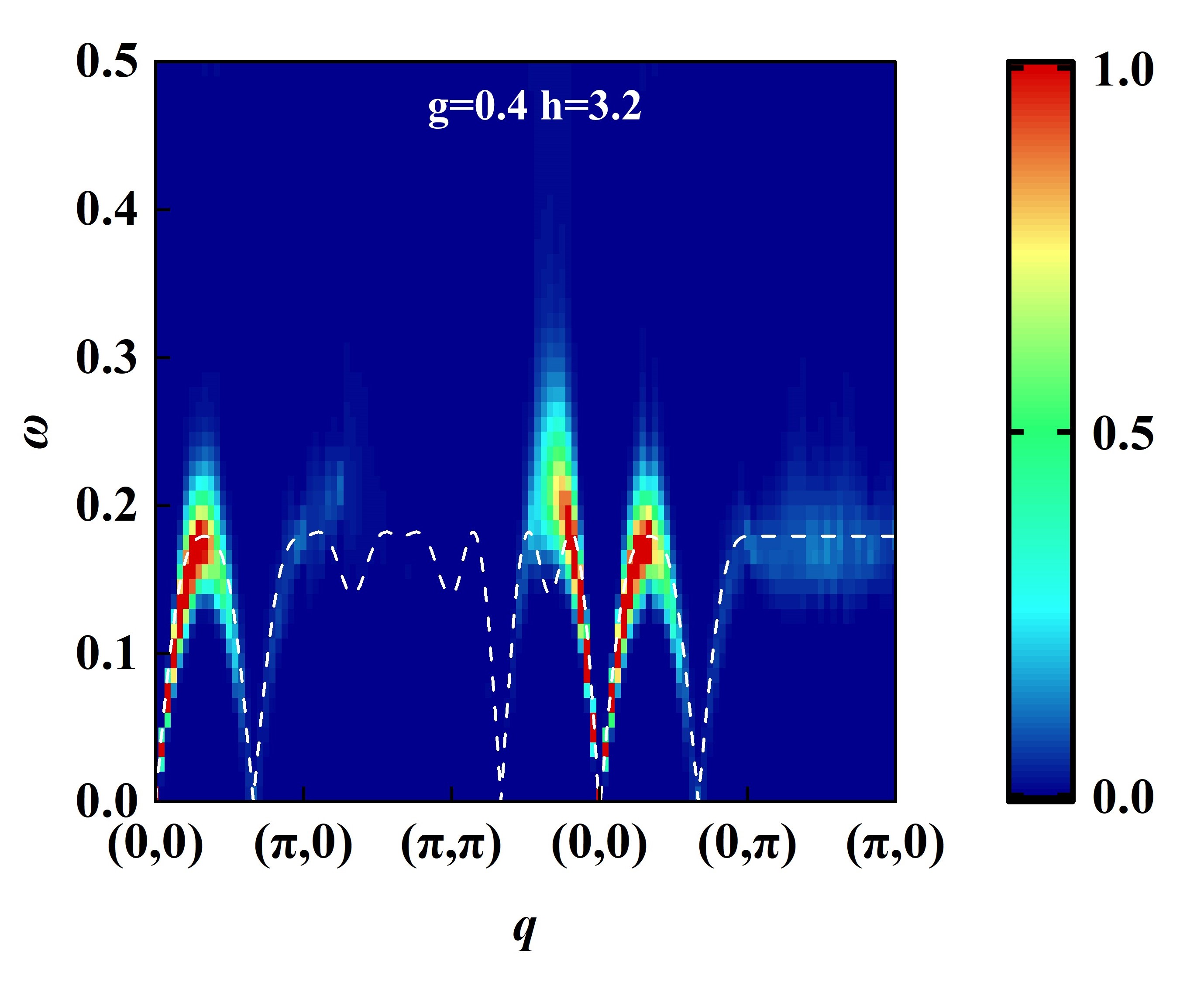}
    \caption{Low-energy gapless branch of g=0.4 and $h$=3.2 extracted from Fig.~\ref{fig:g=0.7and0.4zz}(h). The dashed line is the result of fitting the $\omega$ corresponding to the strongest weight of each wave vector from the low-energy gapless branch with the spin wave of the effective Hamiltonian. The high decay of magnons causes the weight of the spectrum near $(\pi, \pi)$ tend to vanish in the strong magnetic field.}
    \label{fig:g=0.4h=3.2}
\end{figure}

When g=0.4, for the spectra of $S^{zz}(q,\omega )$, the low-energy branch along the path from (0,0) $\to$ $(\pi, 0)$ $\to$ $(\pi, \pi)$ in the weak magnetic field as shown in Fig.~\ref{fig:g=0.7and0.4zz}(e), due to the Brillouin zone folding. As the magnetic field increases, the gap at $(\pi,\pi)$ increases, and the branch with the folding feature merges with high-energy parts and then forms a single magnon excitation mode. In Figs~\ref{fig:g=0.7and0.4zz}(g)-(h), the spontaneous decay of magnons is intenser than in the case of g=0.7. As g decreases, the increase in magnetization under the same magnetic field as shown in Fig.~\ref{fig:phasediagram}, leading to a decrease in the weight of the spectrum. Notably, we observe the distinct separation of the gapped high-energy part and gapless branch in the strong magnetic field as shown in Fig.~\ref{fig:g=0.7and0.4zz}(h). The high-energy part has energy around $J_1$, which originates from intra-sublattice. The low-energy gapless branch exhibits periodic structure and energy around $J_2$, with the gapless points at $(2\pi/3, 0)$ and $(2\pi/3, 2\pi/3)$, due to the folding of the Brillouin zone. To analyze low-energy magnon, we use spin waves of the effective Hamiltonian to fit the $\omega$ of each wave vector with the strongest weight extracted from the low-energy gapless branch, the spin wave is defined as~\cite{spinwave1,spinwave2}

\begin{equation}
\begin{array}{l}
\omega_k=4J_{eff}S_{eff}\sqrt{(1-\gamma_k)(1+\gamma_kcos2\theta)},  
\end{array}
\label{Eq:spinwave}
\end{equation}

where $J_{eff}$ is the effective interaction between 3$\times$3 sublattice, $S_{eff}$ is the effective spin number, $\gamma_k$=($cos{3k}_x$+$cos{3k}_y$)/2 and sin$\theta$=$h_{eff}$/(8$J_{eff}S_{eff}$). We find that the dotted line represents the result of the spin waves of effective Hamiltonian fits the gapless branch well. More details of fitting results can be found in Fig.~\ref{fig:g=0.4h=3.2}. In the spectra of $S^{\pm}(q,\omega )$, Fig.~\ref{fig:g=0.7and0.4xx}(f) show the gapless point at $(\pi,\pi/3)$ and $(\pi/3,\pi/3)$, which originates from the Brillouin zone folding. We observe that the high-energy part with the gap separates from the gapless branch in the strong magnetic field. Fig.~\ref{fig:g=0.7and0.4xx}(i) shows the periodic structure, due to the Brillouin zone folding. Moreover, the gapless branch with obvious characteristics of antiferromagnetic XY phase comes from the excitation of inter-sublattice. In the fully polarized phase, energy gaps appear at the Brillouin zone folding points. Notably, the completely separated low-energy branch is a localized excitation similar to the excitation of the magnetization plateau phase, which we will discuss in detail later.

\begin{figure*}[t]
    \centering
    \includegraphics[width=1\linewidth]{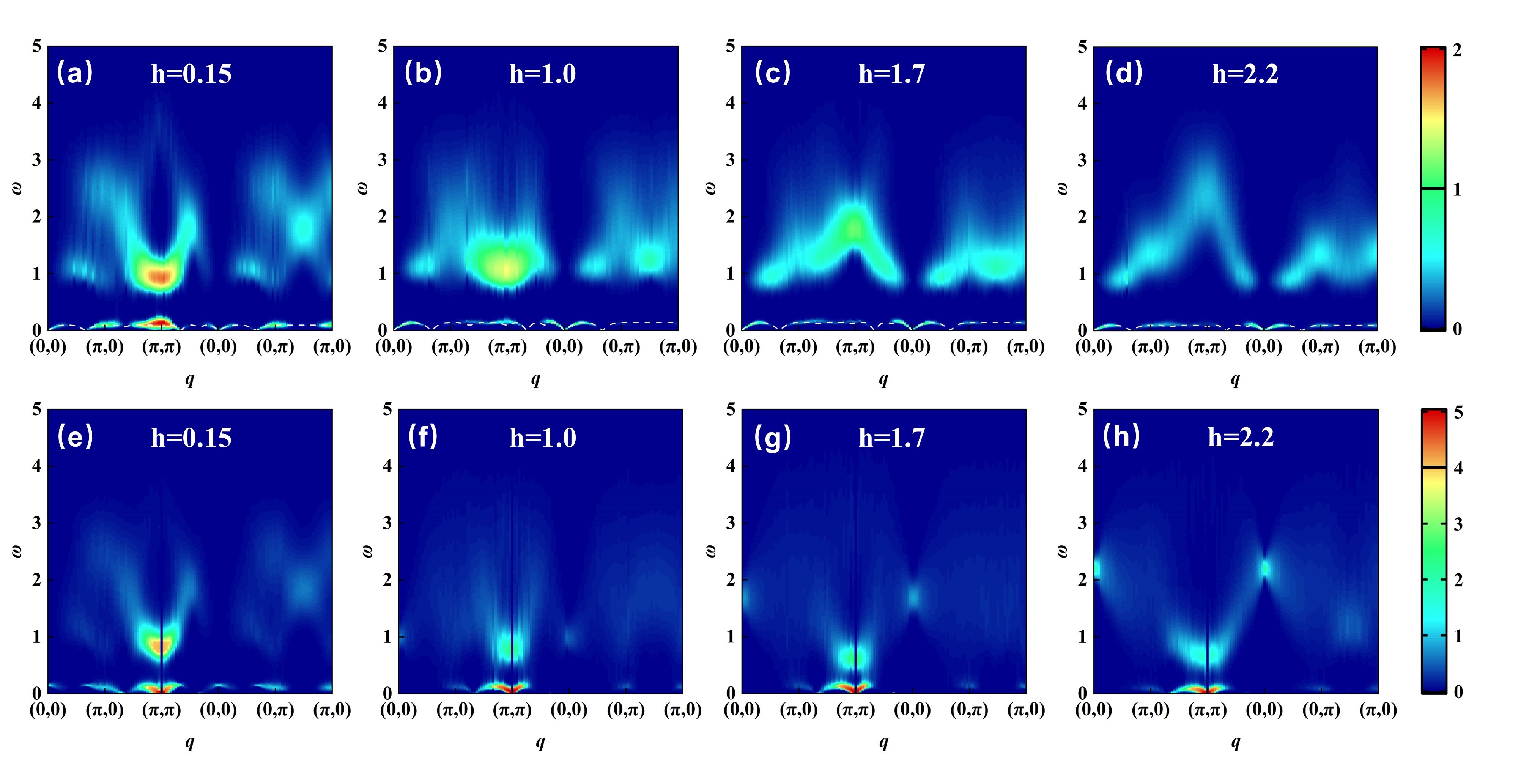}
    \caption{Dynamic spin structure factor $S^{zz}(q,\omega)$ (a)-(d) and $S^{\pm}(q,\omega)$ (e)-(h) of $3\times3$ checkerboard model obtained from QMC-SAC calculations in a fixed coupling ratio g=0.1 with different $h$. Panels (a)-(h) are in the non-plateau XY phase. The dashed lines represent the results of spin wave fitting. We take the values of U$_0$ equal to 1 and 4 for $S^{zz}(q,\omega)$ and $S^{\pm}(q,\omega)$ respectively, corresponding to the black bar as the boundary. Below the boundary, the spectral function is linearly mapped to the color bar, while above the boundary, it is a logarithmic mapping, U = U$_0$ + log$_{10}$$S(q,\omega)$ - log$_{10}$U$_0$.}
    \label{fig:g=0.1zzxx}
\end{figure*}

When g=0.1, we firstly focus on the spectra of $S^{zz}(q,\omega )$ and $S^{\pm}(q,\omega )$ in the non-plateau XY phase, the gapped high-energy and low-energy gapless branch are completely separated even in the weak magnetic field as shown in Fig.~\ref{fig:g=0.1zzxx}. The structure of the gapless branch is similar to Figs.~\ref{fig:g=0.7and0.4zz}(h) and ~\ref{fig:g=0.7and0.4xx}(i). Notably, the values of $J_{eff}S_{eff}$ obtained via spin wave fitting are distinct in different magnetic fields as shown in Table~\ref{tab:SJnihe}, which may be related to the magnetic order of the system. The spectral weight of $S^{zz}(q,\omega )$ and $S^{\pm}(q,\omega )$ are concentrated at the $\omega$ corresponding near the value of magnetic field  at $(\pi, \pi)$ and (0,0), respectively. In Figs.~\ref{fig:g=0.1zzxx} (a) and (e), the spectral weight is concentrated at the low-energy branch, while for other cases is at the high-energy part. At the high-energy regime, the continuum is observed, which is dominated by the excitation of intra-sublattice.

\begin{figure*}
    \centering
    \includegraphics[width=1\linewidth]{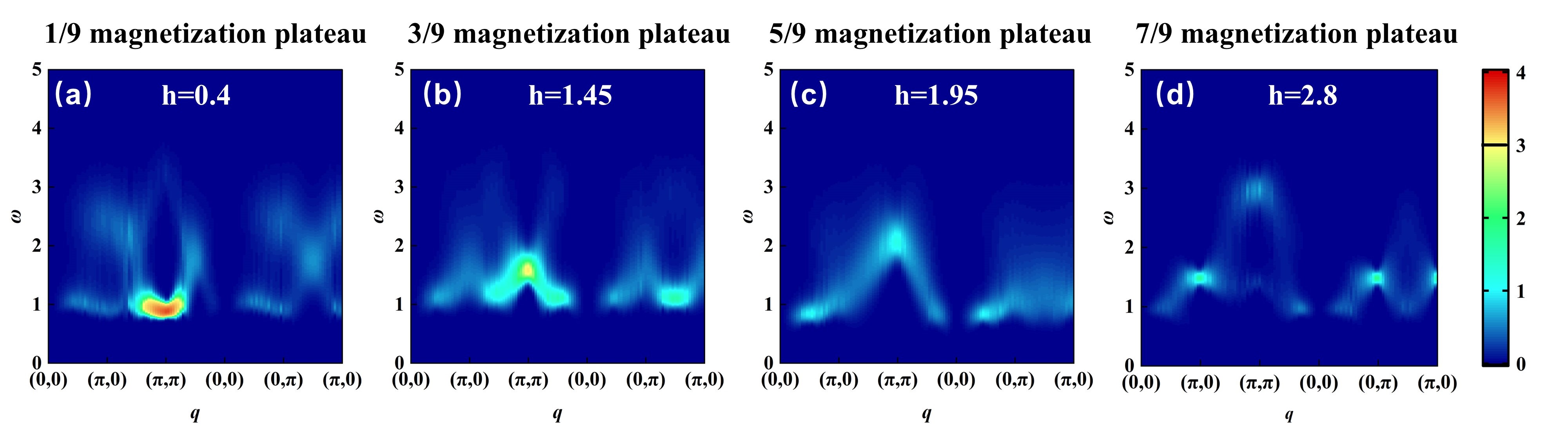}
    \caption{Dynamic spin structure factor $S^{zz}(q,\omega)$ of $3\times3$ checkerboard model obtained from QMC-SAC calculations in a fixed coupling ratio g=0.1 in the different magnetization plateau phases. We take the value U$_0$=3 corresponding to the black bar as the boundary. Below the boundary, the spectral function is linearly mapped to the color bar, while above the boundary, it is a logarithmic mapping, U = U$_0$ + log$_{10}$$S(q,\omega)$ - log$_{10}$U$_0$.}
    \label{fig:plateauzz}
\end{figure*}

\begin{table}[]
\caption{When g=0.1, the values of $J_{eff}$$S_{eff}$ are obtained by fitting the $\omega$ corresponding to the strongest weight of each wave vector from the low-energy gapless branch in different magnetic fields with the spin wave of the effective Hamiltonian.}
\begin{tabularx}{0.4\textwidth}{c c}

\hline\hline
Magnetic field $h$    \qquad\qquad\qquad\qquad\qquad    & $J_{eff}$$S_{eff}$        \\ \hline
0.15                \qquad\qquad\qquad\qquad\qquad        & 0.0239\\
1.0                 \qquad\qquad\qquad\qquad\qquad        & 0.0346\\
1.7                 \qquad\qquad\qquad\qquad\qquad        & 0.035\\
2.2                 \qquad\qquad\qquad\qquad\qquad        & 0.0235\\ 
3.04                \qquad\qquad\qquad\qquad\qquad        & 0.0054\\ \hline\hline
\label{tab:SJnihe}
\end{tabularx}
\end{table}

\begin{figure*}
    \centering
    \includegraphics[width=1\linewidth]{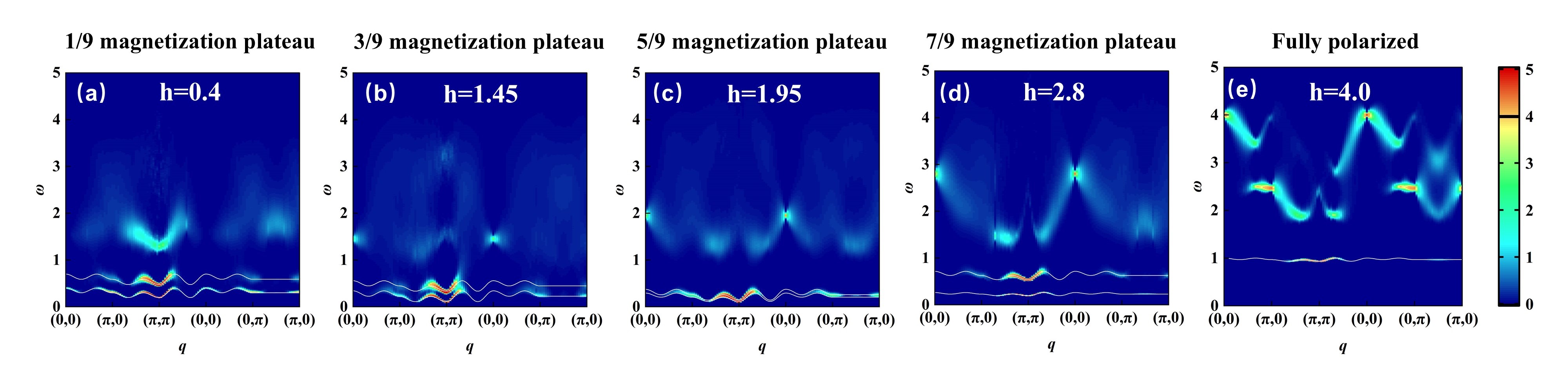}
    \caption{Dynamic spin structure factor $S^{\pm}(q,\omega)$ of $3\times3$ checkerboard model obtained from QMC-SAC calculations in a fixed coupling ratio g=0.1 in different magnetization plateau phase. The white solid lines represent the dispersion relations of $|\Delta m|$=1. We take the value U$_0$=4 corresponding to the black bar as the boundary. Below the boundary, the spectral function is linearly mapped to the color bar, while above the boundary, it is a logarithmic mapping, U = U$_0$ + log$_{10}$$S(q,\omega)$ - log$_{10}$U$_0$.}
    \label{fig:plateauxx}
\end{figure*}

In the magnetization plateau phase and fully polarized phase, we take $h$ as 0.5, 1.45, 1.95, and 2.8 respectively corresponding to 1/9, 3/9, 5/9, and 7/9 magnetization plateau to study the dynamic properties $S^{zz}(q,\omega)$ and $S^{\pm}(q,\omega)$ as shown in Figs.~\ref{fig:plateauzz} and ~\ref{fig:plateauxx}. We observe that the low-energy gapless excitations vanish, and the excitations are gapped in the all momentum space as expected and have energy around $J_1$ in the magnetization plateau phase, which is dominated by the localized excitation from the intra-sublattice. For the $S^{zz}(q,\omega)$, the spectrum structures in respective magnetization plateaus are invariant for the same g. Notably, for the $S^{\pm}(q,\omega)$, we find the lower spectrum of low-energy excitations with $\Delta$m=-1 transition to the high-energy regime, while the high-energy excitations with $\Delta$m=1 transition to the low-energy regime. The weight of the spectrum is concentrated at the low-energy branches near $(\pi, \pi)$. It should be noted that these excitations are internal excitations of $3\times3$ sublattice propagating in momentum space. To explain the excitation mechanism of the energy branches that transition with the magnetic field, we adopt the perturbation theory, which are discussed in detail later. The results of perturbation theory are agree well with QMC-SAC as shown in Fig.~\ref{fig:plateauxx}(a)-(e). In the fully polarized phase $h$=4.0, the energy gap of the Brillouin zone folding point increases, and the high-energy magnon exhibits a periodic structure. The flat bands originate from localized excitation, which matches the results of perturbation theory.

\section{Discussion}
\label{Sec:Discussion}
\begin{figure}
    \centering
    \includegraphics[width=1\linewidth]{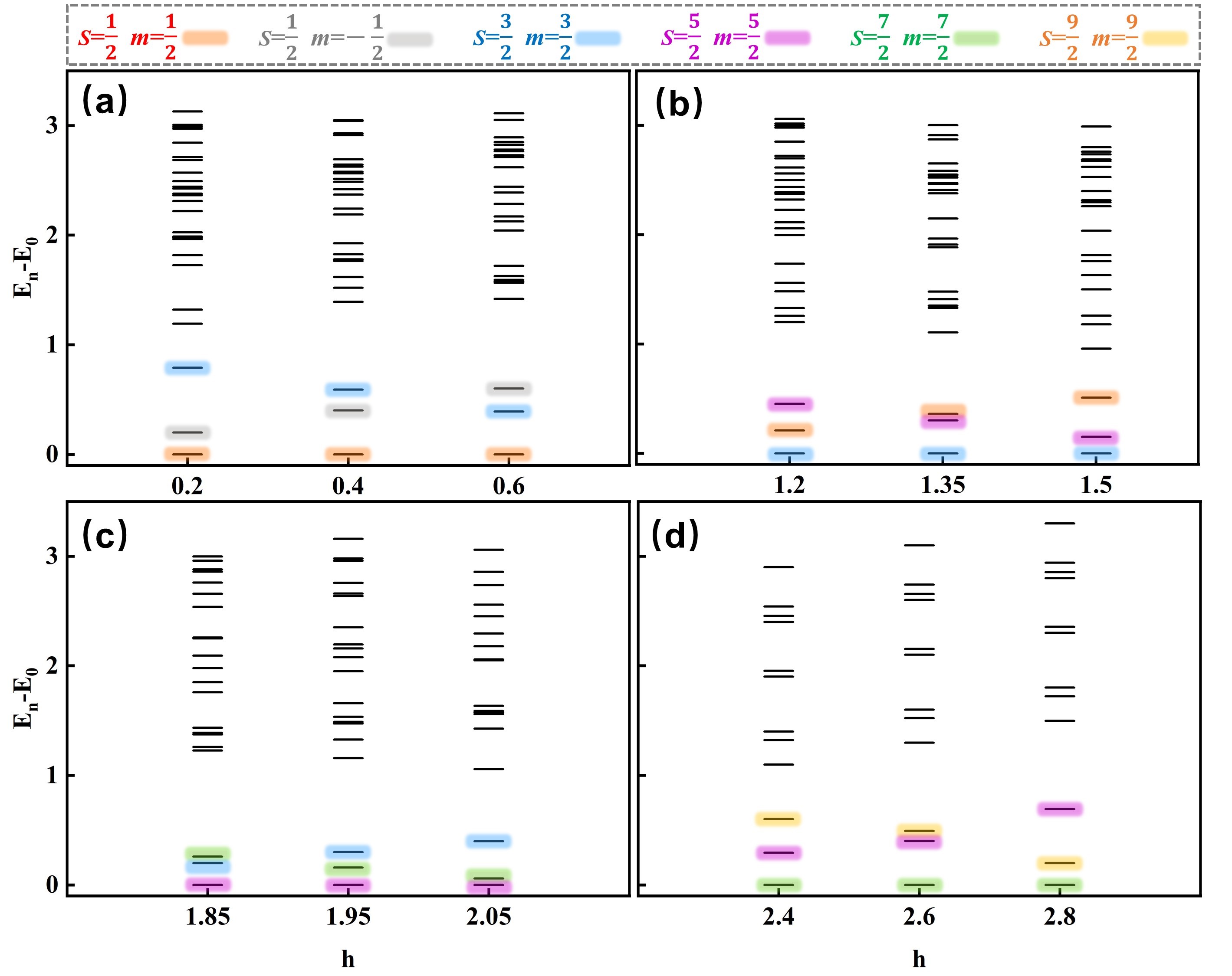}
    \caption{Energy levels with $|\Delta m|=1$ of $3\times3$ isolated plaquette in different phases, where (a)1/9 magnetization plateau phase, (b)3/9 magnetization plateau phase, (c)5/9 magnetization plateau phase and (d)7/9 magnetization plateau phase.}
    \label{fig:EDdeltam=1}
\end{figure}

When g is small, the system emerges the magnetization plateau phase as the magnetic field increases, and the $J_2$ can be considered as a perturbation of isolated $3\times3$ sublattice. To better understand the excitation in the magnetization plateaus, it is meaningful to analyze the excitation spectrum of the isolated $3\times3$ sublattice. We use exact diagonalization to analyze the excitation spectrum of isolated $3\times3$ sublattice in a longitudinal magnetic field. The magnetic field induces the energy level splitting, while the spin quantum numbers and magnetic quantum numbers remain invariant. In Fig~\ref{fig:EDdeltam=1}, the ground state of the 1/9, 3/9, 5/9 and 7/9 magnetization plateaus are S=1/2 m=1/2, S=3/2 m=3/2, S=5/2 m=5/2 and S=7/2 m=7/2, respectively. For low-energy excited states, the energy levels with $\Delta$m=-1 transition to the high-energy part, while energy levels with $\Delta$m=1 transition to the low-energy regime and become the ground state of the next magnetization plateau until the system enters the fully polarized phase. 

To investigate the excitation mechanism of the $3\times3$ checkerboard model in small coupling $J_2$, we adopt a perturbation analysis. In the magnetization plateau phase and fully polarized phase, the magnetic quantum numbers of each $3\times3$ sublattice are 1/2, 3/2, 5/2, 7/2, and 9/2 corresponding the 1/9, 3/9, 5/9, 7/9 magnetization plateaus and fully polarized phase. Therefore, we can construct the wave function of this model by multiplying the ground state of each isolated $3\times3$ sublattice

\begin{equation}
\begin{array}{l}
\left | \psi_g  \right \rangle=\left | 0 \right \rangle _1\otimes\left | 0 \right \rangle _2\cdots\left | 0 \right \rangle _r\cdots\left | 0 \right \rangle _N,
\end{array}
\label{Eq:weirao}
\end{equation}
where  $\left | 0 \right \rangle$ is the ground state of isolated $3\times3$ sublattice. If the r-th sublattice is excited from ground state $\left | 0 \right \rangle$ to $\left | n \right \rangle$ excited state, the wave function of the model is

\begin{equation}
\begin{array}{l}
\left | \psi_e \right \rangle=\left | 0 \right \rangle _1\otimes\left | 0 \right \rangle _2\cdots\left | n \right \rangle _r\cdots\left | 0 \right \rangle _N,
\end{array}
\label{Eq:weirao}
\end{equation}
the excited state in momentum space can be obtained through the Fourier transform
\begin{equation}
\begin{array}{l}
\left | \psi^q  \right \rangle =\frac{1}{\sqrt{N} }  {\textstyle \sum\limits_{r=1}^{N}} e^{-iqr} \left | \psi  \right \rangle_r.
\end{array}
\label{Eq:weirao}
\end{equation}
Thus, the dispersion relations in the reduced Brillouin zone can be obtained as

\begin{equation}
\begin{array}{l}
\epsilon(q)=\left\langle\psi_e^q\right| H\left | \psi_e^q  \right \rangle- \left\langle\psi_g^q\right| H\left | \psi_g^q  \right \rangle.
\end{array}
\label{Eq:weirao}
\end{equation}

According to the Hamiltonian, we find that the result of dispersion relations are mainly dependent on the excited $3\times3$ sublattice and their neighbors. For the low-energy part of the transverse excitation spectra as shown in Fig.~\ref{fig:plateauxx}, the dispersion relations of low-energy excitation from $\left | 0  \right \rangle$ to $\left | n  \right \rangle$ with $|\Delta m|$=1 can well describe the QMC results of the low-energy excitation with strong spectral weight no matter how large the magnetic field is. However, due to the results of QMC-SAC showing the continuum in the high-energy region in the magnetization plateau phase, it is challenging to distinguish the corresponding dispersions.

\section{CONCLUSION}
\label{Sec:CONCLUSION}
In this paper, we use the SSE-QMC to investigate the phase diagram of the $S$=1/2 Heisenberg model with a 3$\times$3 checkerboard structure in a longitudinal magnetic field. The phase diagram presents the non-plateau XY phase, 1/9, 3/9, 5/9, and 7/9 magnetization plateau phases, and the fully polarized phase. We obtain the precise phase transition critical point via finite-size scalings and verify that the phase transition belongs to the 3D XY universality class. The model exhibits 1/9, 3/9, 5/9, and 7/9 magnetization plateaus with magnetic field for small g. The width of the magnetization plateaus will decrease until it completely vanishes under the competition of interaction and magnetic field as the g increases. Notably, the magnetizations and spin correlations remain invariant in the magnetization plateau phase.

Furthermore, the longitudinal and transverse dynamic structural factors exhibit different properties in a longitudinal magnetic field. In the non-plateau XY phase, gapless Goldstone modes are observed at (0,0) for $S^{zz}(q,\omega)$ and $(\pi,\pi)$ for $S^{\pm}(q,\omega)$. When g is large enough, the structures of the spectrum are similar to the square lattice in a magnetic field. As g decreases, the intense magnetic field causes the continuum to split into low-energy gapless branches and high-energy parts, where the low-energy gapless branches with periodic structure can be described by the spin wave in the canted antiferromagnetic phase of effectively Hamiltonian. In the magnetization plateau phase, the excitations are gapped and localized, where the perturbation theory captures the low-energy localized excitations well in the transverse spectra. In the fully polarized phase, the spectra are composed of magnon and localized high-energy excitations for small g. Notably, previous works have successfully developed a viable method for implementing the expected checkerboard model in the optical lattice through cold atom experiments~\cite{opticallattice1,opticallattice2,opticallattice3,opticallattice4,opticallattice5}. This work helps us better understand the properties of quantum magnetic materials with the structure of this model in a longitudinal magnetic field and provides a theoretical basis for checkerboard models with more odd spins sublattice in a longitudinal magnetic field.

\begin{acknowledgments}
We thank Anders W. Sandvik, Han-Qing Wu, Jun-Qing Cheng, Zenan Liu, and Muwei Wu for helpful discussions. This project is supported by NKRDPC-2022YFA1402802, NSFC-92165204, Leading Talent Program of Guangdong Special Projects (201626003), Guangdong Provincial Key Laboratory of Magnetoelectric Physics and Devices (No. 2022B1212010008), and Shenzhen Institute for Quantum Science and Engineering (No. SIQSE202102).

\end{acknowledgments}

\bibliography{mainNotes}

\appendix

\end{document}